\documentclass[showpacs,preprintnumbers,superscriptaddress]{revtex4}
\usepackage{CJK}
\usepackage{amsmath,amssymb,graphicx,bm}
\begin{document}

\title{Thin accretion disks around a black hole in Einstein-Aether-scalar theory}

\author{Tong-Yu He}
\affiliation{College of Physics Science and Technology, Hebei University, Baoding 071002, China}

\author{Ziqiang Cai}
\affiliation{College of Physics Science and Technology, Hebei University, Baoding 071002, China}

\author{Rong-Jia Yang \footnote{Corresponding author}}
\email{yangrongjia@tsinghua.org.cn}
\affiliation{College of Physics Science and Technology, Hebei University, Baoding 071002, China}
\affiliation{Hebei Key Lab of Optic-Electronic Information and Materials, Hebei University, Baoding 071002, China}
\affiliation{National-Local Joint Engineering Laboratory of New Energy Photoelectric Devices, Hebei University, Baoding 071002, China}
\affiliation{Key Laboratory of High-pricision Computation and Application of Quantum Field Theory of Hebei Province, Hebei University, Baoding 071002, China}

\begin{abstract}
We consider the accretion process in the thin disk around a supermassive black hole in Einstein-aether-scalar theory. We probe the effects of the model parameter on the physical properties of the disk. The results show that with increasing value of the parameter, the energy flux, the radiation temperature, the spectra cut-off frequency, the spectra luminosity, and the conversion efficiency of the disk decrease. The disk is hotter and more luminous than that in general relativity for negative parameter, while it is cooler and less luminous for positive parameter. We also find some values of the parameter allowed by the theory are excluded by the physical properties of the disk.
\end{abstract}

\maketitle

\section{Introduction}

To quest for a theory of quantum gravity, one usual way is to introduce a new fundamental symmetry such as supersymmetry, or to look for breaking
a fundamental symmetry such as Lorentz symmetry (LS). The LS is an exact symmetry in special relativity and quantum field theory, while in general relativity (GR) it is only a local symmetry in local inertial frames. The breaking of LS might occur at the Planck or quantum gravity scale. In matter interactions the LS is highly constrained by some precision experiments, while the violation of LS in gravitational sector is not as well explored. Jacobson and his collaborators proposed a LS breaking gravitational theory, called Einstein-Aether theory (EAT), by introduced a unitary time-like vector field, known as aether, in the gravitational action integral \cite{Jacobson:2000xp, Eling:2003rd, Jacobson:2007veq}. The existence of an aether field will lead to a selection of a preferred frame, meaning there is a violation of the LS \cite{Heinicke:2005bp}. The quadratic quantities of the kinematic terms of the aether field involve no more than two derivatives, which lead to a second-order theory as the case of GR. The EAT also describes the classical limit of Ho\v{r}ava gravity \cite{Jacobson:2013xta}. Because of these advantages, it attracts a great deal of attention as a modified gravity theory.

Black holes (BHs) are the ideal objects for studying modified gravity theories \cite{Psaltis:2008bb}. Some interesting black hole solutions in EAT were investigated in literature. For vacuum case, exact BH solutions in EAT were presented in \cite{Eling:2006ec}. BHs with parameterized post-Newtonian parameters identical to those of GR were determined in \cite{Tamaki:2007kz}. In $n$-dimensional spacetime, charged BH solutions with or without the cosmological constant term were investigated in \cite{Lin:2017cmn}. Some rotating BH solutions in EAT were given in \cite{Barausse:2015frm, Ding:2019mal}. Recently, exact solutions in a static spherically symmetric background space with a spacelike aether field were found in \cite{Dimakis:2020dqs}.

The study of accretion disks around BHs is one of the possible methods for detecting the difference between GR and alternative theories. The accretion disk is formed by the diffuse material in orbital motion around a central compact body. The steady-state thin accretion disk is the simplest theoretical model in which the disk has negligible thickness \cite{Shakura1973A, Thorne:1974ve, Page:1974he}. The physical properties of matter forming a thin accretion disk in a variety of background spacetimes have been discussed extensively, see for example \cite{Harko:2009rp, Chen:2011wb, Liu:2021yev, Heydari-Fard:2021ljh, Karimov:2018whx, Chen:2011rx, Kazempour:2022asl, Guzman:2005bs, Gyulchev:2021dvt, Heydari-Fard:2020ugv, Liu:2020vkh, Liu:2019mls, Pun:2008ua, Gyulchev:2020cvo, Gyulchev:2019tvk, Zhang:2021hit}. The signatures appeared in the energy flux and the spectrum emitted by the disk can provide us not only the information about BHs, but also the verification of modified gravity theories \cite{Bambi:2015kza}. Therefore, the study on properties of the thin disk around BHs in EAT could help us to probe the effects of LS breaking in future astronomical observations.

The goal of this paper is to study the properties of the thin accretion disk around a BH in Einstein-aether-scalar theory (EACT) and see whether it can leave us the signature of LS breaking in the energy flux and the spectrum emitted in the mass accretion process. The outline of this paper is organized as follows. In Sec. II, we will briefly review BH solutions in EACT. In Sec. III, we will present the geodesic equations for a particle in the spacetime of BH in EACT. In Sec. IV, we will investigate the physical properties of thin accretion disk around a BH in EACT. Finally, we will briefly summarize our results in Sec. V.

\section{Black hole solutions in Einstein-aether-scalar theory}
In this section, we will briefly review BH solutions in EACT. The gravitational action for the EACT is given by \cite{Kanno:2006ty}
\begin{eqnarray}
\label{1}
S=\int d^4x \sqrt{-g}\left[\frac{R}{2}-\frac{1}{2}g^{\mu\nu}\nabla_{\mu}\phi\nabla_{\nu}\phi-V(\phi)+L_{\ae}\right],
\end{eqnarray}
where $R$ is the Ricci scalar, $g$ is the determinant of the four dimensional metric of the space-time with signatures $(-, +, +, +)$. Here we use the units which fix the speed of light and the gravitational constant: $8\pi G=c=1$. The Lagrangiean density of the aether field is in the following form
\begin{eqnarray}
\label{2}
L_{\ae}=-M^{\alpha\beta}_{~~~\mu\nu}\nabla_{\alpha}u^{\mu}\nabla_{\beta}u^{\nu}+\lambda_0(u^{\mu}u_{\mu}+\varepsilon),
\end{eqnarray}
where $\nabla_\mu$ denotes the covariant derivative with respect to $g_{\mu\nu}$, $\lambda_0$ is a Lagrange multiplier, and $\varepsilon=\pm 1, 0$ is the constant which serves to fix the value of the velocity of the aether field as $u^{\alpha}u_{\alpha}=-\varepsilon$, and. While $M^{\alpha\beta}_{~~~\mu\nu}$ is defined as
\begin{eqnarray}
\label{3}
M^{\alpha\beta}_{~~~\mu\nu}=\beta_1(\phi)g^{\alpha\beta}g_{\mu\nu}+\beta_2(\phi)\delta^{\alpha}_{\mu}\delta^{\beta}_{\nu}
+\beta_3(\phi)\delta^{\alpha}_{\nu}\delta^{\beta}_{\mu}+\beta_4(\phi)u^{\alpha}u^{\beta}g_{\mu\nu},
\end{eqnarray}
where the functions $\beta_1(\phi), \beta_2(\phi), \beta_3(\phi)$ and $\beta_4(\phi)$ define the coupling between the gravitational field and the aether field. Defining $J^{\alpha}_{\mu}=M^{\alpha\beta}_{~~~\mu\nu}\nabla_{\beta}u^{\nu}$, the field equations for the metric take the form
\begin{eqnarray}
\label{4}
R_{\mu\nu}-\frac{1}{2}g_{\mu\nu}R=T^{\ae}_{\mu\nu}+T^{\rm{scalar}}_{\mu\nu},
\end{eqnarray}
with
\begin{eqnarray}
\label{5}
T^{\rm{scalar}}_{\mu\nu}=\nabla_{\mu}\phi\nabla_{\nu}\phi-\frac{1}{2}g_{\mu\nu}\left[\nabla_{\alpha}\phi\nabla^{\alpha}\phi+2V(\phi)\right],
\end{eqnarray}
and
\begin{eqnarray}
\label{6}
T^{\ae}_{\mu\nu}=2\beta_1(\phi)(\nabla_{\mu}u^{\alpha}\nabla_{\nu}u_{\alpha}-\nabla_{\alpha}u_{\mu}\nabla^{\alpha}u_{\nu})
-2\left[\nabla_{\alpha}\left(u_{(\mu}J^{\alpha}_{\nu)}\right)+\nabla_{\alpha}\left(u^{\alpha}J_{(\mu\nu)}\right)
-\nabla_{\alpha}\left(u_{(\mu}J_{\nu)}^{~\alpha}\right)\right]\\\nonumber
-2\beta_4u^{\alpha}u^{\beta}\nabla_{\alpha}u_{\mu}\nabla_{\beta}u_{\nu}+g_{\mu\nu}L_{\ae}
+2\left[u_{\beta}\nabla_{\alpha}J^{\alpha\beta}+\beta_4u^{\alpha}u^{\beta}\nabla_{\alpha}u_{\kappa}\nabla_{\beta}u^{\kappa}\right]u_{\mu}u_{\nu},
\end{eqnarray}
the energy momentum tensors for the scalar and the aether fields respectively. Additionally there also exist the equation
of motion for the aether field $u^{\mu}$, which is
\begin{eqnarray}
\label{7}
\nabla_{\mu}J^{\mu\nu}+\beta_4u^{\kappa}\nabla_{\kappa}u_{\lambda}\nabla^{\nu}u^{\lambda}=\lambda_{0}u^{\nu},
\end{eqnarray}
and for the scalar field $\phi$ that leads to
\begin{eqnarray}
\label{8}
\nabla_{\mu}\nabla^{\mu}\phi-\frac{dV}{d\phi}-\frac{d\beta_1}{d\phi}\nabla^{\mu}u^{\nu}\nabla_{\mu}u_{\nu}
-\frac{d\beta_2}{d\phi}(\nabla^{\mu}u_{\mu})^2-\frac{d\beta_3}{d\phi}\nabla^{\mu}u^{\nu}\nabla_{\nu}u_{\mu}
+\frac{d\beta_4}{d\phi}u^{\alpha}u^{\beta}\nabla_{\alpha}u_{\kappa}\nabla_{\beta}u^{\kappa}=0.
\end{eqnarray}
If the aether field has an additional property, $u^{\alpha}\nabla_{\alpha}u_{\kappa}=0$, then the function $\beta_4$ becomes irrelevant since all its contributions are trivial. A space-like aether field ($\varepsilon=-1$) considered in \cite{Dimakis:2020dqs} fulfils this additional property, and a static spherically symmetric solution with a cosmological constant $\Lambda$ has been derived \cite{Dimakis:2020dqs}
\begin{eqnarray}
\label{9}
\mathrm{d}s^{2}=-\left ( 1-\frac{2\kappa }{r^{2\mu +1} } +lr^{2}  \right ) \mathrm{d}t^{2}+\frac{e^{C}r^{4\mu }\mathrm{d}r^{2}}{1-\frac{2\kappa }{r^{2\mu +1}} +lr^{2}} +r^{2}\left(\mathrm{d}\theta^{2}+\sin^{2}\theta\mathrm{d} \varphi^{2}\right),
\end{eqnarray}
where $l= -\frac{2\mu +1}{2\mu +3} \Lambda$, and the corresponding scalar field is
\begin{eqnarray}
\label{101}
\phi(r)=\pm\frac{e^{\frac{C}{2}}r^{2\mu}}{\sqrt{-\mu(2\mu+1)}}.
\end{eqnarray}
For $\Lambda=0$, the restrictions on $\mu$ are $\mu>-1/2$ and $\mu\neq 0$ to obtain a EACT (since for $\mu= 0$, the EACT reduces to the general relativity and the equation (\ref{9}) reduces to the Schwarzschild metric). For $\mu>0$, the scalar field $\phi$ is imaginary and behaves like a phantom field in the action. On the contrary, we have a canonical scalar field for $-1/2<\mu<0$. For $\Lambda\neq 0$, the restrictions on $\mu$ is $\mu> 0$. We can see that $\kappa$ is associated with the mass of the BH: $\kappa\sim M^{1+2\mu}$ in the units under considering. The line element carries an additional constant $C$ that emerges from the matter content. Therefore this is a hairy BH since another constant appears in addition to the mass.

\section{The geodesic equations}
In this section, we will calculate the equations of motion and the effective potential for a particle in the spacetime (\ref{9}) in EACT. The Lagrangian $\mathcal{L}$ for a point particle around the BH is given by
\begin{eqnarray}
\label{11}
\mathcal{L}=\frac{1}{2}g_{\mu\nu}\frac{\mathrm{d}x^{\mu}}{\mathrm{d}s}\frac{\mathrm{d}x^{\nu}}{\mathrm{d}s}=\frac{1}{2}\varepsilon,
\end{eqnarray}
where $\varepsilon=0$ represents the photon and $\varepsilon=1$ should be considered for a massive particle. We only consider the orbits in the equatorial plane $\theta=\pi/2$. The conserved energy $E$ and the conserved angular momentum $L_{\rm{z}}$ can be calculated, respectively, as
\begin{eqnarray}
\label{12}
E=-g_{tt}\frac{\mathrm{d}t}{\mathrm{d}s}=\left ( 1-\frac{2\kappa }{r^{2\mu +1} } +lr^{2}  \right )\frac{\mathrm{d}t}{\mathrm{d}s},
\end{eqnarray}
%%%%%%%%%%%%%%%%%%%%
\begin{eqnarray}
\label{13}
L_{\rm{z}} =g_{\varphi\varphi}\frac{\mathrm{d}\varphi}{\mathrm{d}s}=r^{2}\frac{\mathrm{d}\varphi}{\mathrm{d}s}.
\end{eqnarray}
The geodesic equations for a massive particle in the spacetime (\ref{9}) take the following forms
\begin{eqnarray}
\label{14}
\left(\frac{\mathrm{d}r}{\mathrm{d}s}\right)^{2}=\frac{1}{e^{C}r^{4\mu }} \left [ E^{2} - \left(1-\frac{2\kappa }{r^{2\mu +1} } +lr^{2}\right)\left(1+\frac{L^{2}_{z}}{r^{2}}\right)  \right ],
\end{eqnarray}
%%%%%%%%%%%%%%%
\begin{eqnarray}
\label{15}
\left(\frac{\mathrm{d}r}{\mathrm{d}\varphi}\right)^{2}=\frac{1}{L^{2}e^{C}r^{4\left ( \mu-1 \right) }}\left [ E^{2} - \left(1-\frac{2\kappa }{r^{2\mu +1} } +lr^{2}\right)\left(1+\frac{L^{2}_{z}}{r^{2}}\right)  \right ],
\end{eqnarray}
%%%%%%%%%%%%%%%%%%%%%5
\begin{eqnarray}
\label{16}
\left(\frac{\mathrm{d}r}{\mathrm{d}t}\right)^{2}=\frac{1}{E^{2}}\left[ 1-\frac{2\kappa }{r^{2\mu +1} } +lr^{2}\right]^{2}\left [ E^{2} - \left(1-\frac{2\kappa }{r^{2\mu +1} } +lr^{2}\right)\left(1+\frac{L^{2}_{z}}{r^{2}}\right)  \right ].
\end{eqnarray}
Note that we can describe the dynamics of the system completely by using Eqs. (\ref{14}-\ref{16}). From Eq.(\ref{14}), the effective gravitational potential can be obtained as
\begin{eqnarray}
\label{17}
V_{\rm{eff}}=-g_{\rm{tt}}\left(1+\frac{L^{2}_{z}}{r^{2}}\right)
=\left(1-\frac{2\kappa }{r^{2\mu +1} } +lr^{2}\right)\left(1+\frac{L^{2}_{z}}{r^{2}}\right).
\end{eqnarray}
Introducing the dimensionless quantities
\begin{eqnarray}
\label{18}
\bar{r}=r\kappa^{-\frac{1}{2\mu +1}},~~~~\bar{l}=l\kappa^{\frac{2}{2\mu+1}},~~~~\bar{L}_{z}=L_{z}\kappa^{-\frac{2}{2\mu+1}},
\end{eqnarray}
%%%%%%%%%%%%%%%%%%%%%%%%
the effective potential can be rewritten as
\begin{eqnarray}
\label{19}
V_{\rm{eff}}=\left(1-\frac{2}{\bar{r}^{2\mu +1}} +\bar{l}\bar{r}^{2}\right)\left(1+\frac{\bar{L}^{2}_{z}}{\bar{r}^{2}}\right).
\end{eqnarray}
For stable circular orbits in the equatorial plane, we have the following conditions: $dr/ds=0$ and
$V_{\rm{eff},r}=0$, where the comma denotes a derivative with respect to the radial coordinate $r$. The angular velocity, the specific energy and the specific angular momentum of particles moving in circular orbits in the spacetime (\ref{9}) are found to be
\begin{eqnarray}
\label{20}
\Omega=\frac{d\varphi}{dt}=\sqrt{\frac{-g_{tt,r}}{g_{\varphi\varphi,r}}}=\sqrt{\frac{\kappa \left ( 2\mu +1 \right ) }{r^{2\mu +3}}+l}
=\frac{\sqrt{\frac{2\mu +1}{\bar{r}^{2\mu +1}}+\bar{l}\bar{r}^{2}}}{\kappa^{\frac{1}{1+2\mu}}\bar{r}},
\end{eqnarray}
%%%%%%%%%%%%%%%
\begin{eqnarray}
\label{21}
E=\frac{-g_{tt}}{\sqrt{-g_{tt}-g_{\varphi\varphi}\Omega^{2}}}=\frac{1-\frac{2\kappa }{r^{2\mu +1} } +lr^{2}}{\sqrt{1-\frac{\kappa \left ( 2\mu +3 \right ) }{r^{2\mu +1}}} }=\frac{1-\frac{2}{\bar{r}^{2\mu +1}} +\bar{l}\bar{r}^{2}}{\sqrt{1-\frac{2\mu +3 }{\bar{r}^{2\mu +1}}} },
\end{eqnarray}
%%%%%%%%%%%%%%%%%
\begin{eqnarray}
\label{22}
L_{\rm{z}}=\frac{g_{\varphi\varphi} \Omega}{\sqrt{-g_{tt}-g_{\varphi\varphi}\Omega^{2}}}=\frac{\sqrt{\frac{\kappa \left ( 2\mu +1 \right )}{r^{2\mu -1}} +lr^{4} }  }{\sqrt{1-\frac{\kappa \left ( 2\mu +3 \right ) }{r^{2\mu +1}}}}=\frac{\kappa^{\frac{1}{1+2\mu}}\bar{r}\sqrt{\frac{2\mu +1}{\bar{r}^{2\mu +1}}+\bar{l}\bar{r}^{2}}}{\sqrt{1-\frac{ 2\mu +3 }{\bar{r}^{2\mu +1}}}}.
\end{eqnarray}
For a test particle in the gravitational potential of a central body, the innermost stable circular orbit, known as the ISCO radius, is given by the condition: $V_{\rm{eff},rr}=0$ or $V_{\rm{eff},\bar{r}\bar{r}}=0$. For $r<r_{\rm{isco}}$, the equatorial circular orbits are unstable, so $r_{\rm{isco}}$ determines the inner edge of the thin accretion disk.

\section{Physical properties of thin accretion disks}
In this section, we first briefly review the physical properties of thin accretion disks based on the Novikov-Thorne model \cite{Novikov:1973kta} which is a generalization of the Shakura-Sunyaev model \cite{Shakura1973A}. The model has some typical assumptions: (1) the space-time of the central massive object is stationary, axisymmetric and asymptotically flat; (2) the disk's mass has no effect on the background metric; (3) the accretion disk is geometrically thin, namely its vertical size is negligible, compared to its horizontal size; (4) the orbiting particles around the compact central object move between the outer edge $r_{\rm{out}}$ and $r_{\rm{isco}}$ which determines the inner edge of the disk; (5) the accretion disk lies in the equatorial plane of the accreting compact object, namely the disk surface is perpendicular to the BH spin; (6) resulting from hydrodynamic and thermodynamic equilibrium of the disk, the emitted electromagnetic radiation from the disk surface is assumed to have a black body spectrum; (7) the mass accretion rate of the disk, $ \dot{M_{0}}$, does not change with time.

In the following subsections, we will investigate in detail about the radiant energy flux, the radiation temperature, the observed luminosity, and some important values related to the thin accretion disk. Even with numerical calculation, it is difficult to find solutions to the equation $V_{\rm{eff},\bar{r}\bar{r}}=0$. Second, the spacetime in Novikov-Thorne model needs to be asymptotically flat. For simplicity but without loss generality, we focus on the zero cosmological constant case.

We use the following values for physical constants and the thin accretion disk: $c=2.997\times 10^{10}$ cms$^{-1}$, $\dot{M}_{0}=2\times10^{-6} M_{\odot}$yr$^{-1}$, 1yr$=3.156\times10^7$s, $\sigma_{\rm{SB}}=5.67\times 10^{-5}$erg s$^{-1}$cm$^{-2}$K$^{-4}$, $h=6.625\times 10^{-27}$ergs, $k_{\rm{B}}=1.38\times 10^{-16}$ ergK$^{-1}$, $M_{\odot}=1.989\times10^{33}$g, and the mass of BH $M=2\times 10^6M_{\odot}$. Because theoretically it is impossible to give the value of $e^C$ in the metric (\ref{9}) and there is yet no research on constraining it by using observational data, we choose $e^C$ to be equal to $-4\mu$ power of the gravitational radius of a sun $e^C=(1.48\times 10^{5}{\rm{cm}})^{-4\mu}$.

\begin{figure}
\centering
\includegraphics[height=2in,width=3in]{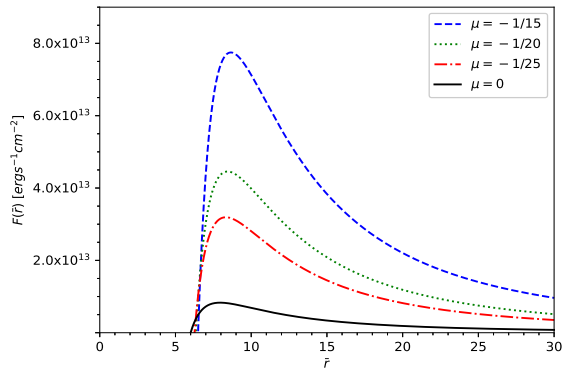}
\caption{The energy flux $F(\bar{r})$ from a disk around a BH in EACT for different negative values of $\mu$.}
\label{fig1}
\end{figure}

\begin{figure}
\centering
\includegraphics[height=2in,width=3in]{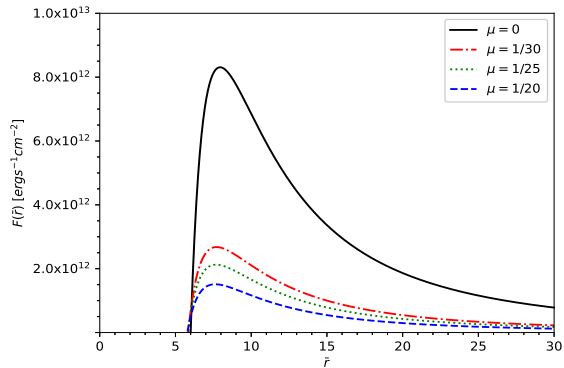}
\caption{The energy flux $F(\bar{r})$ from a disk around a BH in EACT for different positive values of $\mu$.}
\label{fig2}
\end{figure}

\subsection{The radiant energy flux}
		
From the conservation equations of rest mass, energy, and the angular momentum of the disk particles, the radiant energy flux over the disk surface can be obtained as \cite{Novikov:1973kta, Page:1974he}
%%%%%%%%%%%%%%%%%%5	
\begin{eqnarray}
\label{230}
F(r)=-\frac{\dot{M}_{0} \Omega_{, r}}{4 \pi \sqrt{-g}(E-\Omega L_{\rm{z}})^{2}} \int_{r_{\mathrm{isco}}}^{r}(E-\Omega L_{\rm{z}}) L_{\rm{z},r} \mathrm{d}r.
\end{eqnarray}
This equation was widely used in the literature. It is valid only for cylindrical coordinates. If adopting spherical coordinates, it takes the form \cite{Collodel:2021gxu}
\begin{eqnarray}
\label{231}
F(r)=-\frac{\dot{M}_{0} \Omega_{, r}}{4 \pi \sqrt{-g/g_{\theta\theta}}(E-\Omega L_{\rm{z}})^{2}} \int_{r_{\mathrm{isco}}}^{r}(E-\Omega L_{\rm{z}}) L_{\rm{z},r} \mathrm{d}r.
\end{eqnarray}
For the case considered here, we rewrite it in terms of dimensionless parameter $\bar{r}$ as
\begin{eqnarray}
\label{23}
F(r)=-\frac{\dot{M}_{0} \Omega_{, \bar{r}}\kappa^{-\frac{1}{1+2u}}}{4 \pi \sqrt{-g/g_{\theta\theta}}(E-\Omega L_{\rm{z}})^{2}} \int_{\bar{r}_{\mathrm{isco}}}^{\bar{r}}(E-\Omega L_{\rm{z}}) L_{\rm{z},\bar{r}} \mathrm{d}\bar{r}.
\end{eqnarray}
Here the mass accretion rate, $\dot{M}_0$, whose value we consider is in the range for supermassive BHs, see for example SgrA*, a supermassive BH at the center of the Milky Way with a mass of order $M = 4.1\times10^6M_{\odot}$ and with an estimated rate $\dot{M}_{0}=10^{-9} -10^{-7}M_{\odot}$yr$^{-1}$.

Figures \ref{fig1} and \ref{fig2} show the energy flux $F(\bar{r})$ of a disk around a BH in EACT for different values of $\mu$. We see that whether for positive or negative value of $\mu$, the energy flux decreases as the value of $\mu$ increases. For negative $\mu$, the deviation from Schwarzschild BH ($\mu=0$) decreases as the value of $\mu$ increases; while for positive $\mu$, the deviation increases as the value of $\mu$ increases. The values of the energy flux for negative $\mu$ is larger than those for positive $\mu$. This property may help to use astronomical observations to constrain the parameter $\mu$.

\subsection{The radiation temperature}
\begin{figure}
\centering
\includegraphics[height=2in,width=3in]{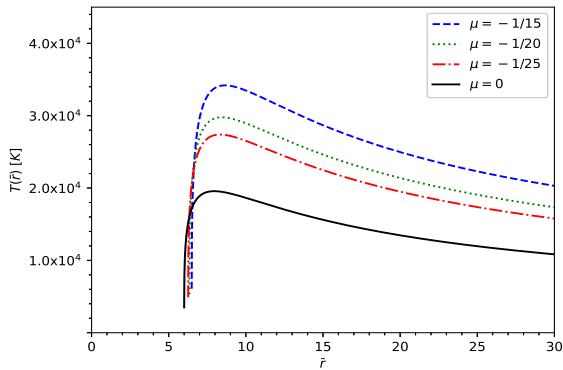}
\caption{Variety of the disk temperature $T(\bar{r})$ with negative parameters $\mu$ for the thin disk around a BH in EACT.}
\label{fig3}
\end{figure}

\begin{figure}
\centering
\includegraphics[height=2in,width=3in]{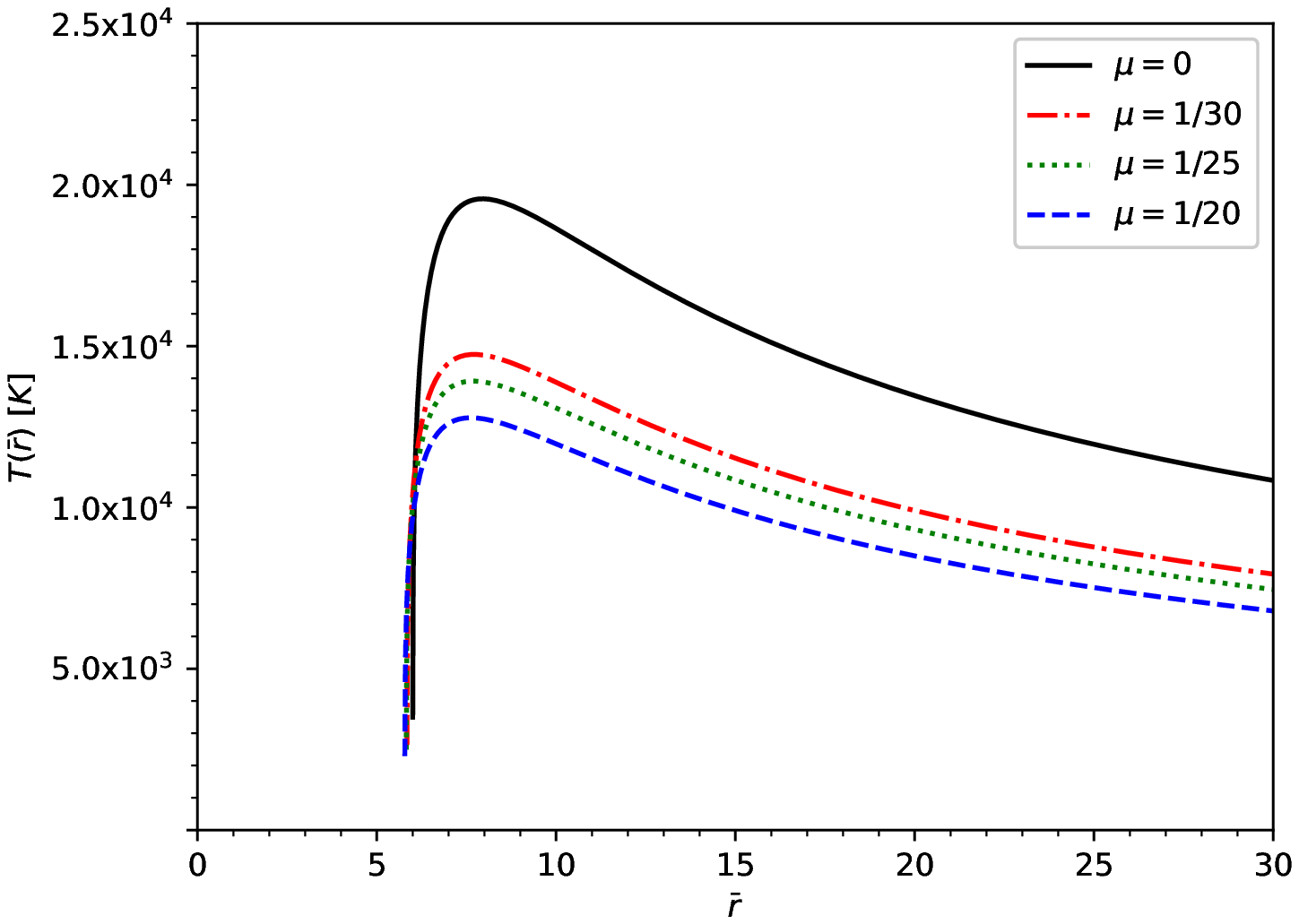}
\caption{Variety of the disk temperature $T(\bar{r})$ with positive parameters $\mu$ for the thin disk around a BH in EACT.}
\label{fig4}
\end{figure}

In the Novikov-Thorne model, the accreted matter is in thermodynamic equilibrium, meaning the radiation emitted by the disk can be considered as a perfect black body radiation. The radiation temperature $T(r)$ of the disk is related to the energy flux $F(r)$ via the Stefan-Boltzmann law, $F(r)=\sigma_{\rm{SB}}T^4(r)$, where $\sigma_{\rm{SB}}$ is the Stefan-Boltzmann constant. This indicates that the dependence of $T(\bar{r})$ on $\bar{r}$ is similar to that of the energy flux $F(\bar{r})$ on $\bar{r}$, as shown in figures \ref{fig3} and \ref{fig4}. With increasing negative $\mu$, the disk temperature decreases and the maximum values shift closer to the inner edge of the disk; with increasing positive $\mu$, the disk temperature decreases and the maximum values also shift closer to the inner edge of the
disk. Moreover, for negative values of $\mu$, the disk around a BH is much hotter than the disk around a Schwarzschild BH ($\mu=0$) in GR, while for positive values it is cooler and less efficient.
%%%%%%%%%%%%%%%%%%%%5
%%%%%%%%%%%%%%%%%%%%%%%%

\subsection{The observed luminosity}
\begin{figure}
\centering
\includegraphics[height=2in,width=3in]{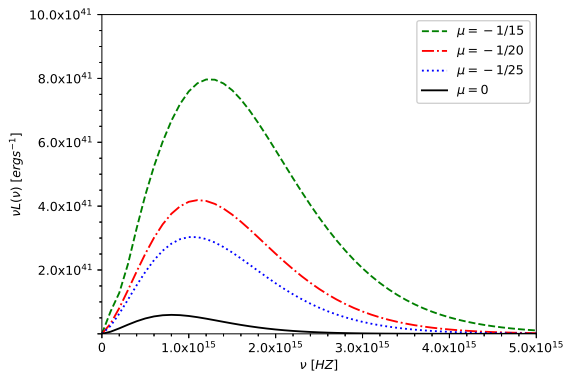}
\caption{The emission spectrum $\nu L(\nu)$ of the accretion disk around a BH in EACT for different negative values of $\mu$, as a function of frequency $\nu$.}
\label{fig5}
\end{figure}

\begin{figure}
\centering
\includegraphics[height=2in,width=3in]{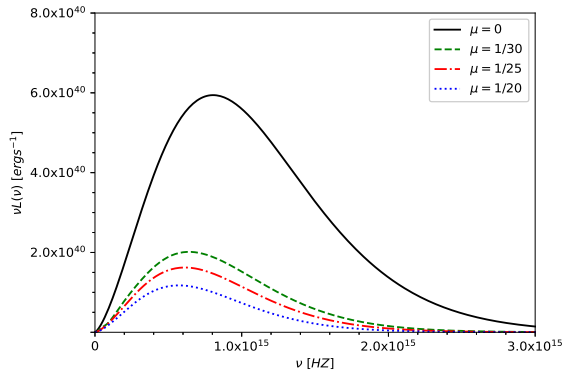}
\caption{The emission spectrum $\nu L(\nu)$ of the accretion disk around a BH in EACT for different positive values of $\mu$, as a function of frequency $\nu$.}
\label{fig6}
\end{figure}

The observed luminosity $L(\nu)$ for the thin accretion disk around a BH has a red-shifted black body spectrum given by \cite{Torres:2002td}
\begin{eqnarray}
\label{24}
L(\nu)=4 \pi \mathrm{d}^{2} I(\nu)=8 \pi h \cos\gamma \int_{r_{\rm{i}}}^{r_{\rm{f}}} \int^{2\pi}_0\frac{\nu^{3}_{e} r drd\varphi}{e^{\frac{h \nu_{e}}{k_{\rm{B}}T}}-1}
=8 \pi h \kappa^{\frac{2}{1+2\mu}}\cos\gamma \int_{\bar{r}_{\rm{i}}}^{\bar{r}_{\rm{f}}} \int^{2\pi}_0\frac{\nu^{3}_{e} \bar{r} d\bar{r}d\varphi}{e^{\frac{h \nu_{e}}{k_{\rm{B}}T}}-1},
\end{eqnarray}
where $d$ is the distance to the disk center, $I(\nu)$ is the thermal energy flux radiated by the disk, $h$ is Planck constant, $k_{\rm{B}}$ is the Boltzmann
constant, and $\gamma$ is the disk inclination angle which we will set to be zero. The quantities $r_{\rm{f}}$ and $r_{\rm{i}}$ are outer and inner radii of the edge of the disk, respectively. Assuming that the flux over the disk surface vanishes at $r_{\rm{f}} \longrightarrow \infty $, we choose $r_{\rm{i}}$ = $r_{\rm{isco}}$ and $r_{\rm{f}} \longrightarrow \infty $ to calculate the luminosity $L(\nu)$ of the disk.  The emitted frequency is given
by $\nu_{e}= \nu(1 + z)$, where the redshift factor can be written as
\begin{eqnarray}
\label{}
1+z=\frac{1+\Omega r\sin\varphi\sin\gamma}{\sqrt{-g_{tt}-2\Omega g_{t\varphi}-\Omega^2g_{\varphi\varphi}}},
\end{eqnarray}
where the light bending is neglected \cite{Luminet:1979nyg, Bhattacharyya:2000kt}.
Applying the Eq. (\ref{23}), we present the spectral energy distribution of the disk radiation in figures \ref{fig5} and \ref{fig6}. Similar
to the case of the energy flux and the disk temperature, we see that for negative $\mu$, the disk around a BH in EACT is more luminous than that around a Schwarzschild BH ($\mu=0$) in GR, while for positive $\mu$ it is less luminous.
%%%%%%%%%%%%%%%%%%%%%%%%%%%%%%%%%%%%%%
%%%%%%%%%%%%%%%%%%%%%%%%%%5

\subsection{The Novikov-Thorne efficiency and some important values}
In the mass accreted process around a BH, the radiative efficiency is an important characteristic quantity which describes the capability of the central object converting rest mass into outgoing radiation. If all the emitted photons can escape to infinity, we can find the efficiency $\epsilon$ is determined by the specific energy of a particle at the marginally stable orbit $r_{\rm{isco}}$ \cite{Novikov:1973kta, Page:1974he}
\begin{eqnarray}
\label{25}
\epsilon=1-E_{\rm{isco}}.
\end{eqnarray}
In Table \ref{tab1}, we show the maximum values of the energy flux $F_{\rm{max}}(\bar{r})$, the disk temperature $T_{\rm{max}}(\bar{r})$, and the $\nu L(\nu)_{\rm{max}}$ of the accretion disk around a BH in EACT. The cut-off frequency $\nu_{\rm{crit}}$ and the Novikov-Thorne efficiency $\epsilon$ are also presented. We see that with increasing value of $\mu$, all these values decrease. In other words, the values of these quantities are larger than those of the accretion disk around a Schwarzschild BH in GR for negative $\mu$; while the opposite is true for positive $\mu$.

\begin{table}
  \centering
  \begin{tabular}{c c c c c c c }
  \hline
  \hline
  $\mu$ & $\bar{r}_{\rm{isco}}$ & $F_{\rm{max}}(\bar{r})$ergs$^{-1}$cm$^{-2}$ & $T_{\rm{max}}(\bar{r})K$ & $\nu_{\rm{crit}}$Hz & $\nu L(\nu)_{\rm{max}}$ergs$^{-1}$ & $\epsilon$\\
  \hline
  $-1/15$ & 6.49 & $7.75\times 10^{13}$ & $3.42\times 10^{4}$ & $1.20\times10^{15}$ & $7.97\times10^{41}$ & 0.081\\
  \hline
  $-1/20$ & 6.34 & $4.46\times 10^{13}$ & $2.98\times 10^{4}$ & $1.10\times10^{15}$ & $4.19\times10^{41}$ & 0.075\\
  \hline
  $-1/25$ & 6.26 & $3.19\times 10^{13}$ & $2.74\times 10^{4}$ &$1.00\times10^{15}$ & $3.03\times10^{41}$& 0.071\\
  \hline
  $0$     & 6 & $8.31\times 10^{12}$ &$1.96\times 10^{4}$ &$8.00\times10^{14}$ &$5.94\times10^{40}$ &0.057 \\
  \hline
  $1/30$ & 5.84 &$2.68\times 10^{12}$ & $1.47\times 10^{4}$ & $6.41\times10^{14}$ &$2.02\times10^{40}$ & 0.047\\
  \hline
  $1/25$ & 5.82 & $2.13\times 10^{12}$ &$1.39\times 10^{4}$ & $6.15\times10^{14}$ &$1.62\times10^{40}$ & 0.046 \\
  \hline
  $1/20$ & 5.78 & $1.51\times 10^{12}$ &$1.28\times 10^{4}$ & $5.74\times10^{14}$ & $1.18\times10^{40}$&0.043 \\
  \hline
  \hline
  \end{tabular}
  \caption{The maximum values of the energy flux $F_{\rm{max}}(\bar{r})$, the disk temperature $T_{\rm{max}}(\bar{r})$, and $\nu L(\nu)_{\rm{max}}$ of the accretion disk around a BH in EACT. The ISCO radius, the cut-off frequency $\nu_{\rm{crit}}$, and the Novikov-Thorne efficiency $\epsilon$ are also shown in the table.}\label{tab1}
\end{table}

In the process of numerical calculation, we found some values of the parameter $\mu$ are allowed by the theory, but do not meet the physical requirements of the accretion disk, see for example, for $\mu=\frac{1}{2}$, the system of equations $V_{\rm{eff},r}=0$ and $V_{\rm{eff},rr}=0$ have no real roots; while for $\mu=-\frac{1}{4}$, some values of the energy flux $F(\bar{r})$ will be imaginary.

The results above show that the parameter $\mu$ related to the coupling between the gravitational field and the aether
field has great impact on the physical properties of thin accretion disk, which may can be used as tests to probe the effects of LS breaking in future astronomical observations.

\section{Conclusions}
In this work, we have studied electromagnetic properties of thin accretion disk around a BH in EACT.
We have calculated the geodesic equations and the effective gravitational potential. We have investigated the effect of the model parameter $\mu$ on the energy flux, the disk temperature, the luminosity spectra, and the radiative efficiency of the disk and showed that with increasing $\mu$, all these quantities decrease. We also have compared the properties of the disk with those around a Schwarzschild BH in GR. We have found that for positive $\mu$ the ISCO radius of disk in EACT is smaller than that of disk around a Schwarzschild BH. This result is to be expected not only because the radius of the black hole, $r_{\rm{H}}=(2\kappa)^{\frac{1}{1+2\mu}}$, in solution (\ref{9}) for $\mu<0$
is larger than that for $\mu>0$, meaning the black hole is more massive for $\mu<0$, but also because the scalar field behaves like phantom for positive $\mu$, weakens the strength of gravity and thus the ISCO radius takes smaller values. For negative $\mu$, the disk around a BH in EACT is hotter and more luminous than that around a Schwarzschild BH in GR, while it is cooler and less luminous for positive $\mu$. We note that the radiative efficiency $\epsilon$ is smaller for smaller values of $\bar{r}_{\rm{isco}}$ for the model considered here, which is in contrast to what happens for the Kerr metric when increasing the spin parameter of the BH \cite{Collodel:2021gxu}. In the numerical calculations, we also have found that some values do not meet the physical requirements of the disk though they are allowed by the theory, see for example, the equations $V_{\rm{eff},r}=0$ and $V_{\rm{eff},rr}=0$ have no real roots for $\mu=\frac{1}{2}$; while some values of the energy flux $F(\bar{r})$ will be imaginary for $\mu=-\frac{1}{4}$. So the disk properties can give physical constraints on the parameter of EACT. All the results show that the parameter $\mu$ has great impact on the physical properties of the disk, which may be used as tests to probe the effects of LS breaking in future observations.

\begin{acknowledgments}
We are very grateful to the anonymous reviewers for their insightful comments, constructive and valuable suggestions. This work is supported in part by Hebei Provincial Natural Science Foundation of China (Grant No. A2021201034).
\end{acknowledgments}

\bibliographystyle{ieeetr}%{elsarticle-num}
\bibliography{acc}
\end{document}